\documentclass[pdflatex,sn-basic]{sn-jnl}

\usepackage{graphicx}
\usepackage{amsmath,amssymb,amsfonts}
\usepackage{booktabs}
\usepackage{caption}
\begin{document}

\title[Risk Assessment]{Criminal Justice Risk Assessments with Unreliable Class Assignments}

\author*{\fnm{Richard} \sur{Berk}}\email{berkr@sas.upenn.edu}
\affil{\orgdiv{Department of Criminology}, \orgname{University of Pennsylvania}, \orgaddress{\city{Philadelphia}, \state{PA}, \country{USA}}}

\abstract{Quantitative risk assessments have been used for decades to help inform criminal justice decisions. A set of predictors and a categorical response variable are used to train a statistical classifier. The trained classifier can be employed to compute risk scores for a new, unlabeled case for which a forecast is needed, one risk score for each response variable class. In the spirit of Bayes classifiers, the class with the largest risk score conventionally becomes the forecasted outcome label for that case. The largest and next largest risk scores can be very similar or very different. However, when they are much the same, the outcome class is essentially being assigned by a coin flip; the classification is unreliable. Forecasting error becomes more likely, and concerns about fairness can arise. This paper shows how Mondrian conformal prediction can increase classification reliability when the classifier is indecisive and provide valid estimates of forecast uncertainty. Risk forecasts for a large dataset of offenders on probation are used to illustrate the issues.}


\maketitle

\section{Introduction}\label{sec1}

Quantitative risk assessments have been used for decades to help inform criminal justice decisions. Illustrations include: determining secure housing placements for prison inmates \citep{Cunningham2010}, inmate release decisions for parole \citep{Burgess1928}, supervision protocols for offenders on probation or parole \citep{Kuchibhotla2023}, judges' sentencing decisions \citep{Berk2014}, sentencing guidelines \citep{Wilkins1978}, ``career criminal'' selective incapacitation \citep{Greenwood1982}, and supervisory options for juvenile offenders \citep{Schwalbe2008}. 

There have been many concerns about risk assessment accuracy. Some well-informed criticisms can be found as early as the 1950s \citep{Reiss1951, Goodman1953a, Goodman1953b}. Other accuracy issues have surfaced far more recently \citep{Berk2014, Berk2023a}. Current criticisms also have addressed fairness \citep{Kleinberg2017} and transparency \citep{Rudin2019}.

Perhaps the broadest rebuttal is a reminder that criminal justice risk assessments by humans are hardly flawless. Deficiencies in all forms of criminal justice risk assessment should berecognized, making the proper benchmark not perfection, but human performance. Quantitative risk assessments just have to do better. This paper proceeds from that point of view.

Classifiers used in risk assessment are often indecisive about particular offenders. The result can be forecasts of outcome classes that are unreliable; they are determined by the rough equivalence of a coin flip. Classifiers by and large neglect this problem and proceed in the spirit of a Bayes classifier \citep{Corbettdavies2017, Berk2017}. The class with the highest classification score is forecasted even if that score is very similar to one or more other scores.
 
This paper offers a solution within an explicit forecasting framework building on Mondrian conformal prediction sets that widen the forecast to multiple outcome classes precisely when the underlying classifier is indecisive \citep{Vovk2003,Sadinle2019}. New kinds of fairness issues then necessarily arise. Unreliability and how to properly address it are illustrated through a project to update a machine learning risk algorithm used by the Department of Probation and Parole in a large American city. 

Section~\ref{sec2} provides some statistical background for the analyses ahead. Section~\ref{sec3} describes the data used in the probation risk assessment construction. Section~\ref{sec4} introduces the statistical procedures applied. Section~\ref{sec5} reports results from the classifier used with the training data. Section~\ref{sec6} provides results for the Mondrian conformal prediction sets. Section~\ref{sec7} offers some policy implications. Section~\ref{sec8} contains a summary and conclusions. The appendix is a didactic discussion of the Mondrian approach to conformal prediction sets.

\section{Some Statistical Background}\label{sec2}

Classifiers for quantitative risk assessment are fit to training data containing categorical response variables such as ``sentenced to prison,'' ``sentenced to jail,'' or ``sentenced to probation.'' Building on the exposition in  Hastie and colleagues (\citeyear{Hastie2009}), $G$ denotes that outcome variable, which takes one of $K$ outcome class labels, $1, 2, 3, \dots, K$. A statistical model or, more recently, a statistical algorithm, is fit to training data containing $G$ and a set of predictors $X$, using one of several possible loss functions that typically are used to minimize the number of misclassifications from the training data. The estimand is expected prediction error: $EPE = E[L(G,\hat{G}(X))]$.\footnote
{
$G$ is the actual outcome class, analogous to $Y$ in a regression problem.
The estimand is $f_{k}(x)$, the true class-conditional function that the
classifier targets (detailed in the next paragraph), analogous to
$E[Y | X = x]$ in regression. The classifier is the estimator,
producing an estimate $\hat{f}_{k}(x)$, from which $\hat{G}$, the
predicted class label, is obtained. There is nothing formal yet about
forecasting despite the word ``prediction'' in ``expected prediction error.''
}
For example, a case in the training data has the label ``sentenced to prison'' whereas the classifier incorrectly assigns the label ``sentenced to jail.''
 
Using the training data, the function estimated by a classifier is denoted by $\hat{f}_{k}(x)$, where $k = 1, 2, \dots, K$. The output is often treated as estimated class probabilities. Some approaches allow each pair of classes to have their own fitting function and fitted values \citep{Lughofer2013}, but this is not common practice.

Classification requires a rule for turning fitted values into an assigned class. In particular, for classifiers that output fitted class probabilities, including random forests used in this paper, the standard rule assigns each case to the class with the highest fitted probability: $\delta(x) = \arg\max_k \hat{f}_k(x)$. This rule targets the true, unknown conditional probability of class membership in the population, $P(G=k \mid X=x)$ \citep{Hastie2009}; $\hat{f}_k(x)$ is the fitted counterpart of that population quantity, estimated from a training sample.\footnote
{
Not all classifiers output probabilities. Support vector machines, for example, classifies using distance from a separating hyperplane rather than a fitted probability \citep{Hastie2009}. This paper is restricted to probability-output classifiers, and random forests specifically.
}

A classifier's performance is customarily evaluated using 0-1 loss: $L(G, \delta(X)) = 0$ if $\delta(X) = G$, and $1$ otherwise, regardless of which incorrect class has been assigned. This is the loss implicit in ``minimizing the number of misclassifications,'' and it is used throughout this paper to evaluate classifier performance; misclassifications are not weighted directly when performance is reported. 

Most fitting procedures do not minimize 0-1 loss directly. A random forest, for instance, grows individual trees by minimizing node impurity, typically Gini impurity or entropy \citep{Breiman2001}. But 0-1 loss remains the standard yardstick once a classifier has been fit. The cost of misclassifying more serious offense classes, for example, can be instead addressed through the training data itself; these classes are oversampled, so that the fitting procedure is pushed to distinguish them more effectively, without altering the loss function used to evaluate performance.

The training product is in practice a forecasting instrument. To help fix these ideas, suppose that a criminal justice agency supervising offenders on probation has a rule that failing to appear for a mandatory drug test while on probation leads to an automatic probation revocation. There is a binary $G$ for failing to appear or not. A classifier is applied to $N$ training observations to produce a classification score (here a probability) for each training data case $i = 1, 2,  \dots, N$. For each case in the training data, the outcome class with the largest probability is the fitted class $\hat{G}$. But because these results are derived from the training, they are largely irrelevant to true forecasting. The goal is to train the algorithm for subsequent steps that produce true forecasts.

Suppose that soon after the training is completed, an offender not in the training data is being considered for discharge from probation. Denote that as the $N + 1$ case. A probation revocation is an undesirable outcome, which, if likely, could lead to a denial of the offender's discharge petition.

Using the same predictor variables employed in training and the already trained algorithm itself, a probability for the $N+1$ case of failing to appear for a mandatory drug test is estimated. Suppose that probability is $0.55$. This is leads to a real forecast; if this inmate is discharged from probation, the estimated probability of failing to appear for a mandatory drug test is $0.55$

What should a criminal justice decision maker do with that probability? It is the larger of the two probabilities (i.e., the probability of complying with the mandatory drug test is 0.45). In the spirit of Bayes risk, the forecast should be non-compliance, which looks like an argument for denying the probation discharge. But that probability means that of each 100 very similar offenders discharged, a little more than half would fail to appear for a mandatory drug test. Would that outcome be much different if each discharge petition was decided by drawing lots? Would that be good agency practice? Would that even be fair? 

In summary, the neat picture of risk assessment classification can stumble when a classifier makes small distinctions between outcome classes. Part of the problem can be omitted variables that might otherwise increase substantially the variation in the risk scores. Also, because the data are composed of random variables, random variation can dilute systematic variation that the included predictors can produce. A principled solution is needed. 

\section{Data}\label{sec3}

The data that support the empirical part of this paper were provided by the criminal justice agency responsible for probation and parole in a large American city. The data span the years 2019 through 2024. More recent data were not available. There are over 70,000 offenders included with over 4000 of the most recent cases having no information yet for the response variable $G$.\footnote
{
There are some parolees as well, but they are a tiny fraction of the total. Although the data technically are available to the public, privacy concerns make them difficult to obtain.
} 

The supervising agency wanted a risk instrument that forecasted a re-arrest within two years of the beginning of a probation or parole sentence.  Agency concerns emphasized re-arrests while under supervision because the vast majority of offenders are under supervision for at least two years.   

Two kinds of re-arrests were defined: a re-arrest for a ``serious'' crime and re-arrest for a crime that was not ``serious.'' Several agency staff members consulted state statutes to decide which kinds of crimes were in which category. Serious crimes were generally violent offenses (e.g., homicide, attempted homicide, armed robbery, rape) and some sex crimes (e.g., involving children). All other crimes were defined as not serious. Highlighting a public safety agenda, the outcome variable classes were ``high risk'' for serious crimes, ``moderate risk'' for crimes that were not serious, and ``low risk'' for the absence of an arrest. One can certainly argue with the content of these categories, but these were the outcome classes that were politically acceptable and potentially forecasted well.\footnote
{
When there were multiple charges, the most serious charge was used to determine the most appropriate outcome class in the training data. Conviction offense was not used because it was seen as some distance from the real offense and a product of processes motivated by other concerns than public safety such as case-by-case strategic negotiations between prosecutors and defense attorneys.  Also, there often can be very long delays between charging and trial, risking a substantial increase in incomplete data. 
}

The forecasts were intended to help determine the intensity and kind of supervision that each offender received. Draconian measures (e.g., house confinement) while under supervision were quite rare. Group cognitive therapy, anger management training, and drug treatment programs were commonly prescribed as well as a range of reintegration programs such as parenting classes, vocational training and educational services. 

The predictors race, zip code of residence, and prior juvenile arrests were excluded. Including this information in a risk analysis had long been political a hot button issue. Exercising an abundance the caution, the agency decided to not to allow their use in instrument construction or later forecasting. They understood that some forecasting accuracy probably would be lost.

Included predictors included the following:

\begin{itemize}
\item age
\item gender
\item recency (i.e., the time since the most recent prior arrest) 
\item the number of prior jail terms
\item the age of the first adult arrest
\item  the number of prior prison terms
\item the number of adult violent priors
\item the number of drug distribution priors
\item the number of adult priors for any crime
\item the number of priors for serious crimes
\item the number of current drug charges
\item the number of current charges for violent crimes
\item the number of current charges for gun crimes
\item the number of current charges for sex crimes
\end{itemize}

The data are treated as random realizations from a common joint probability distribution of local offenders sentenced to probation or parole. Because convicted offenders are sentenced to parole or probation one-by-one by many different judges, it is probably reasonable to treat the realizations as independent. Sensible forecasting generally requires that the generating joint probability distribution does not change in important ways over time. That issue is discussed when Mondrian conformal prediction sets are introduced. Some empirical evidence is brought to bear.

The raw data had 67880 cases with 4131 of those cases having missing data for the response variable. These were most recent cases for which the follow-up data were not yet available. Each of the variables was examined for obvious errors and implausible outliers. Overall, the data provided were very clean but curious outliers were recoded back to reasonable caps (e.g., 50 prior arrest charges and 20 prior jail incarcerations). Prior research had shown that long right tails did not improve forecasting skill (the author) perhaps because they were scattered data entry errors.

Anticipating the use of Mondrian conformal prediction sets and a split sample estimation approach \citep{Sadinle2019, Vovk2003, Angelopoulos2023, Angelopoulos2025}, three randomly split, disjoint subsets were constructed: training data $(N_T = 57880)$, calibration data $(N_C = 5000)$ and forecast data ($N_F = 5000)$ before rows with missing data were deleted. A little more than 5\% of the data overall were lost. The notation for the number of cases after rows with any missing data were deleted is, $N_t$, $N_c$ and $N_f$ respectively.

\section{Statistical Methods}\label{sec4}

The summary statistics and histograms for each predictor in the training data were examined. There were no surprises. For example, the youngest offender in the data is nearly 16 (i.e., tried and convicted as an adult), the oldest offender is nearly 88, the median age is 33, and the interquartile range spans ages 27 to 42. The median number of priors overall (i.e., prior arrest charges) is 35, but the first quartile is 17. There is a large number of first time offenders or offenders with very few priors. This might be expected for offenders sentenced to probation.

It was immediately apparent that the response variable in the training data was highly unbalanced. 9\% of the offenders are high risk, 28\% are moderate risk, and 63\% are low risk. The distribution is consistent with expectations and has important implications for training. 

\subsection{Random Forests as a Classifier}
Random forests \citep{Breiman2001} in \textsf{R}  was used for training. The usual default tuning parameters were maintained. The one exception was the use of stratified bootstrap sampling to make classification error similar across the three outcomes \citep{Hasanin2018}. An important motivation was that there was widespread agreement among stakeholders that misclassifying an offender who is truly high risk as low risk is much worse than the reverse. The same rationale applies to truely, moderate-risk offenders, though less extreme. Oversampling rare but especially important outcome classes when each tree in the forest is grown makes the algorithm work harder to identify those cases. In addition, one can better consider forecasting accuracy, when classification error is very similar across outcome classes. Good forecasting is the main goal of the risk assessment algorithm, and it can be examined unconfounded with classification error that can differ by outcome class.\footnote
{
Classification error is central for fitting the training data, but forecasting error is the policy concern that matters. Classification error treats the outcome class for each case as known and conveys how often the classifier fails to identify it. Prediction error conditions on the fitted class and conveys how often the fitted class is wrong. Prediction error is not forecasting error because it is a product of the training data. But it can provide an initial sense of how well the trained classifier will forecast when cases are unlabeled.} The results of the stratified bootstrap sampling will be apparent when the training data confusion table is discussed.

\subsection{Mondrian Conformal Prediction Sets}
Mondrian conformal prediction will be unfamiliar to many readers, but it is just a variant of the standard split sample approach to conformal prediction regions \citep{Angelopoulos2023, Angelopoulos2025}. A didactic discussion is provided in the appendix. 

 Recall that in addition to the training dataset, there is a calibration dataset and a forecast dataset. All three are constructed as random disjoint subsets of the full dataset. Only the training data with $N_t$ cases were used to fit the random forest. The calibration data and the forecast data remained genuine (``pristine'') holdout samples. All three datasets have the same predictor and response variables. The value of $\alpha$ (i.e., the miscoverage rate) was specified as $0.25$, making specified coverage probability $1-\alpha = 0.75$. Stakeholders chose this coverage after a discussion of the tradeoff between precision and coverage.
 
 The distinctive feature of the Mondrian approach is subsetting the conformity scores constructed from the calibration data by each case's actual outcome class in those data \citep{Sadinle2019, Vovk2003}. If the performance of the classifier depends on the outcome class in the training data, the Mondrian method captures the performance variation in different conformity score distributions. Other split sample approaches, in effect, average over all conformity score distributions. Because in this study, each outcome class has its own collection of consequences, it is important to examine the uncertainty in their prediction sets separately.
 
 Just as with all of the standard conformal inference methods, a particular application of exchangeability is essential. The calibration conformity scores must be exchangeable with the conformity score of each future case. Whether this holds depends on the relationship between the calibration subset and those future cases. This reduces to whether cases arriving after the calibration data were assembled are generated by the same conditional process, given an outcome class, as those responsible for calibration. Ultimately, this is an empirical question. It is addressed here for the offender data through direct evidence of distributional stability over the study window, including the near-replication of confusion table statistics against a risk tool built five years earlier (the author). This evidence bears directly on the stability of the conformity score distribution itself, rather than resting on a structural argument about how the data arose. These claims are supported later by comparing theoretical and empirical coverage across Mondrian prediction sets. If exchangeability does not hold, those prediction sets would likely have substantially lower empirical coverages than the overall specified coverage. 

\section{Training Data Results}\label{sec5}

For a conventional baseline, Table~\ref{tab:All} is the random forest confusion table computed with the OOB (``out-of-bag'') subsets of the training data. The algorithm was tuned so that the classification errors along the table's right margin would be similar. Because classification error assumes that the outcome class is known, it is used primarily to monitor the algorithm's classification performance. For all three outcome classes, when the actual outcome class is known, the random forest correctly identifies it for about two-thirds of the offenders. By this criterion, the algorithm is fitting the data reasonably well. However classification performance does not say much about forecasting performance.\footnote
{ 
If in practice the outcome class were known, there would be no need to forecast it.
}

\begin{table}[htp]
\centering
\begin{tabular}{|c|c|c|c|c|}
\hline \hline
  & Pred. High  &  Pred. Moderate  & Pred. Low  & Class. Error \\
\hline 
Actual High & 3386  &  684 & 833 & 0.31 \\ 
Actual Moderate  & 1798  & 10830  & 2710 &  0.29 \\
Actual Low  & 3965  & 7436 & 22696  &  0.33 \\
\hline
Pred. Error & 0.63 &  0.43  &  \textbf{0.14}  &  \\
\hline
Supervision Percentage & \textbf{17\%} & 35\% & 48\% & N = 54,338 \\
\hline \hline
\end{tabular}
\caption{Confusion Table for All Offenders (Fraction of offenders in each true outcome class within 2 years: High = 0.09, Moderate = 0.28, Low = 0.63)}
\label{tab:All}
\end{table}

The row near the bottom containing prediction error shows the proportion of times that each one of the three \emph{predicted} outcome classes is wrong. Prediction error for high risk  offenders is 63\%, for moderate risk offenders is 43\% and for low risk offenders is 14\%. The random forest algorithm is able to make promising distinctions between offenders who are low risk compared to offenders who are either high or moderate risk. Distinctions between offenders at high or moderate risk are far more difficult to resolve. Moreover, many of the predicted classes are based on small differences in the fitted probabilities making their reliability questionable. For both reasons, prediction performance for high risk and moderate risk is disappointing.

The confusion table also provides information on the manner in which outcome class predictions cascade into parole supervision loads. 17\% of offenders would be supervised initially as high risk, 35\% as moderate risk, and 48\% as low risk. Compared to their representation in the training data shown in the caption, high-risk offenders are substantially ``over-predicted," moderate-risk offenders are somewhat ``over-predicted'' and low-risk offenders are substantially ``under-predicted.'' These differences have significant cost implications. Supervision of projected high- and moderate-risk offenders is far more expensive than supervision of projected low-risk offenders. Generally, low-risk offenders need only keep their contact information current, whereas high- and moderate-risk offenders are subject to far more intensive oversight and are provided with a range of interventions intended to foster reintegration with their families and neighborhoods.

\subsection{Predictor Importance and Functions}

Random forest comes with several functions that provide useful information about the fitting process and results. A common example is variable importance: the size of each predictor's impact on the fitted values. One effective way to assess this is by randomly permuting each predictor in turn and determining how much classification error increases as a result. The larger the increase, the more important the predictor is to the fitted classification. When there are three or more classes, each class will typically show different patterns of predictor importance. Because public safety is a hot-button political issue, variable importance for high-risk offenders is of special interest.

\begin{figure}[h]
    \centering
    \includegraphics[width=0.8\textwidth]{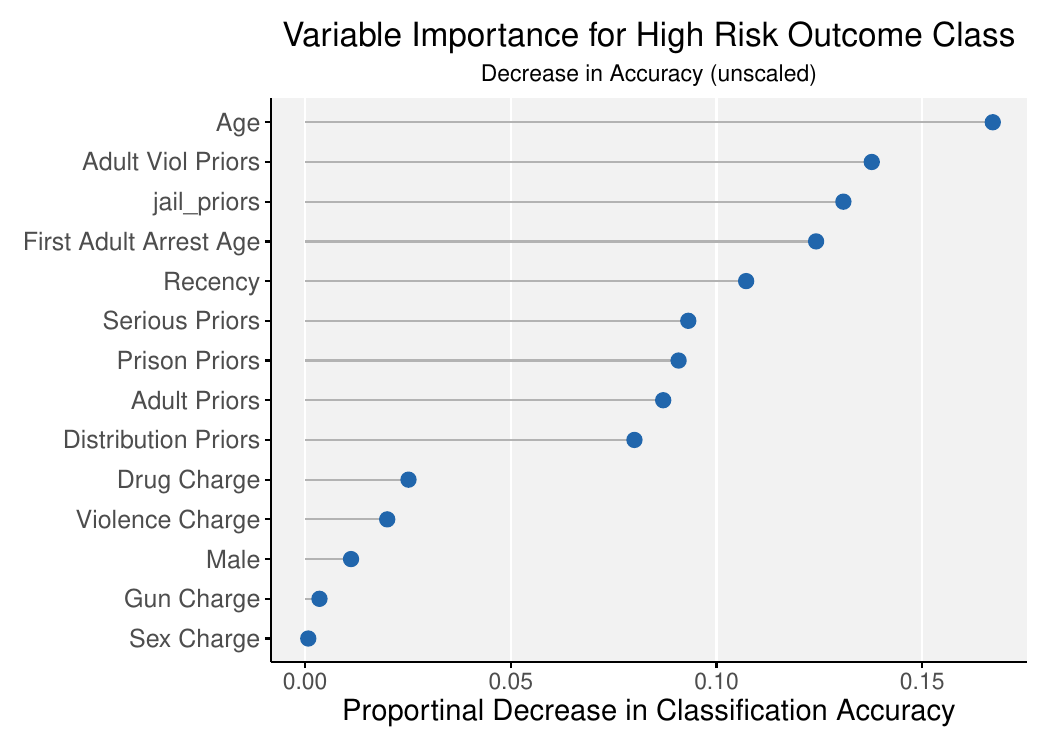}
    \caption{Classification Permutation Variable Importance of High-Risk Outcome Class}
    \label{fig:imp-plot-high}
\end{figure}

Figure~\ref{fig:imp-plot-high} displays the results. The age of the offender when probation begins leads all other predictors, accounting for a decline of over 15\% in classification accuracy when permuted. The next four predictors in order are the number of violent priors as an adult, the number of prior jail terms, the age at which the offender was first arrested as an adult, and the recency of the last arrest before the arrest leading to a sentence of probation. All are responsible  for a decline of more than 10\% in classification accuracy when randomly permuted. Gender hardly matters with all other predictors held constant, nor does whether the current arrest involved serious charges.

\begin{figure}[h]
    \centering
    \includegraphics[width=0.8\textwidth]{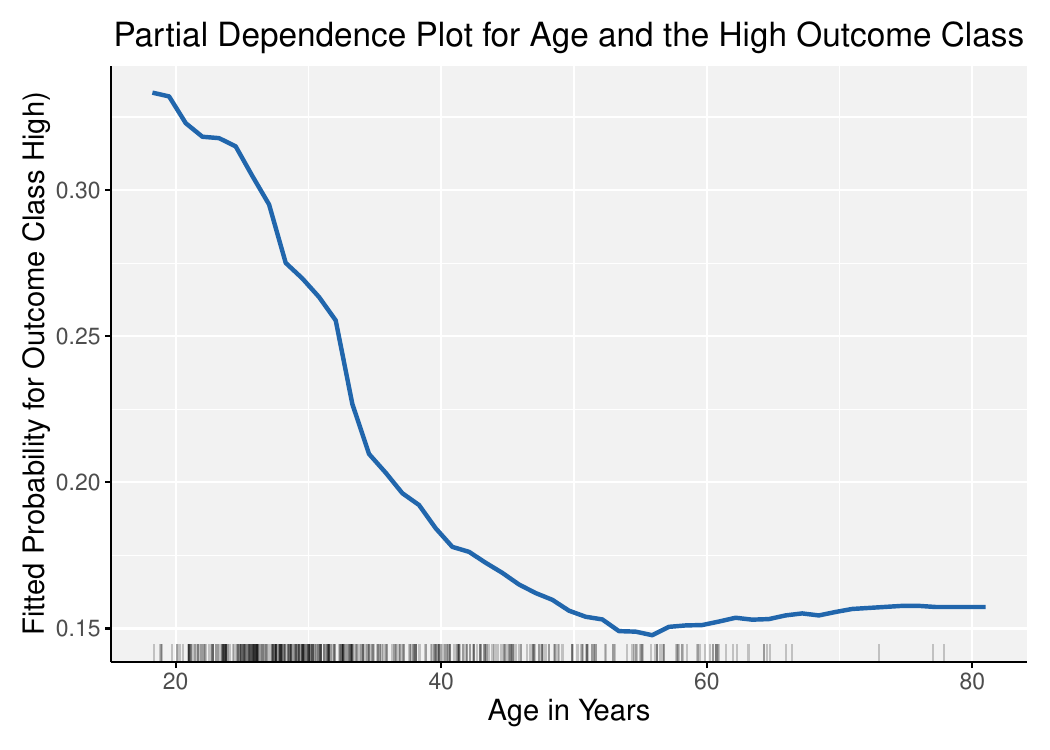}
    \caption{Partial Dependence Plot for Age and the High Risk Outcome Class}
    \label{fig:age-pdp}
\end{figure}

Random forest is not a causal model of any sort. No causal claims are being made about the genesis of crime. But many of the predictors producing these fitted values have long been shown to be related to criminal activity \citep{Gottfredson1987}. The plots serve two purposes: sanity checks for the random forest fit, and answers to questions about what is responsible for a given offender's assigned outcome class. These questions sometimes arise at intake when offenders are told their outcome risk class. They want to know why they were labeled high, moderate, or low risk. Such information makes the classifier's distinctions more transparent.

\begin{figure}[!h]
    \centering
    \includegraphics[width=0.8\textwidth]{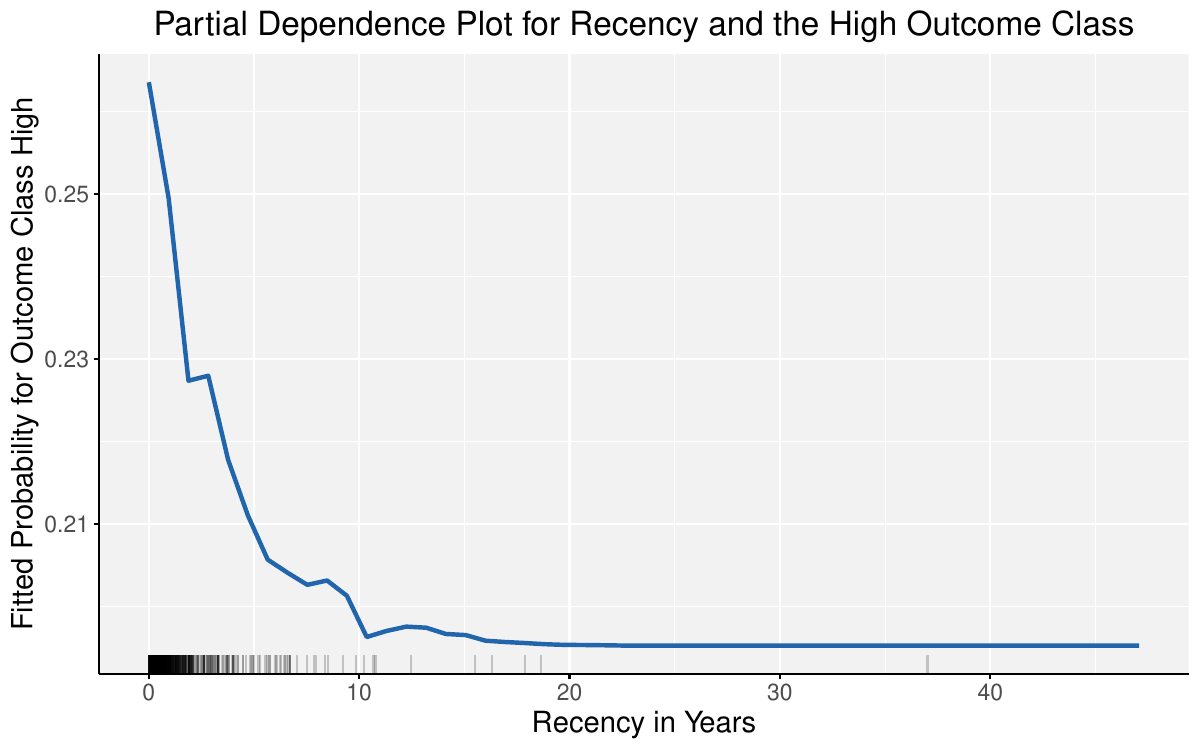}
    \caption{Partial Dependence Plot for the Recency of the Immediately Preceding Arrest and the High-Risk Outcome Class}
        \label{fig:recency-pdp}
\end{figure}

The nature of the relationship between each predictor and the response variable also matters for sanity checks and transparency. Partial dependence plots provide one form this information \citep{Molnar2025}. Consider the predictor age as an example. The computation of partial dependence begins by fixing age at its earliest value in the training data, such as 18, for every offender, replacing the varying ages actually present in the training data. All other predictors retain their original, differing values. This new dataset is passed to the random forest's predict function, producing new fitted values for each offender, one for each outcome class. A single outcome class, such as high risk, is specified, and the average of the new fitted values for that class is computed. In other words, one obtains the average fitted probability of the high-risk outcome class if all offenders were 18, with all other predictor values held at their original levels. The same is done for each year of age, providing a visualization of the conditional probability of the high-risk outcome class for different ages. For computational convenience, a predictor with many values is sometimes first binned. The same steps can be used for any predictor of interest and any outcome class.

Figure~\ref{fig:age-pdp} shows the result for an offender's age at intake from the training data. A rug plot is provided at the bottom of the figure. Almost all of the offenders are younger than 60, with ages most clustered between 20 and 30. The estimated partial dependence function is plotted in blue. It declines dramatically until around age 50 and then levels off. The drop is substantial: from a fitted probability of about 0.35 to a fitted probability of about 0.15. Again, random forest is not a causal model, but one can see that age affects the fitted values approximately as one might expect. This provides another sanity check, as well as a more detailed answer should questions arise about the role of age in how the risk class is assigned.

In a similar fashion, Figure~\ref{fig:recency-pdp} shows the partial dependence plot for recency of the immediately preceding arrest. The mass of the data falls at about 5 years or less. There are relatively few observations beyond 10 years. Just as with age, the fitted probability for the high-risk outcome class falls dramatically and then levels off. The highest probabilities are found at 2 years or less. But the impact on the fitted values is much smaller than for age. The fitted probabilities fall from about 0.26 to 0.20.

Partial dependence plots were constructed for the three other most important predictors, and the results are unsurprising. For example, the greater the number of prior arrests as an adult, the larger the probability of the high-risk outcome class. Here too, the relationship is highly nonlinear. The probability of the high-risk outcome class increases at a decreasing rate and then levels off at about 20 priors. The first several priors matter the most.

\section{Mondrian Conformal Prediction Regions}\label{sec6}

For all variants on conformal prediction sets, a prediction set might include a single outcome such as ``Low'' when there is an unambiguous forecast of low risk. But prediction sets can include more than one outcome class, such as ``High'' and ``Moderate,'' when the risk scores for the two outcome classes are very similar. Usually, prediction sets containing more than one outcome class provide ambiguous information; more than one outcome class is being forecasted. But sometimes such results are instructive nevertheless. 

A discussed earlier, the Mondrian approach subsets the calibration conformity scores into separate distributions, one conformity score distribution for each outcome class. The statistical theory is correct and powerful, but depends on conformity score exchangeability.  In real data, exchangeability is often a matter degree. It can be good practice to use a holdout sample to evaluate whether empirical performance is consistent with exchangeability. It is also important to consider whether the risk assessment forecasts are compatible with agency practices and responsive to political concerns.

Because the forecast holdout sample constructed earlier includes the response variable class for each case, it can be used to empirically evaluate the performance of conformal prediction sets showing what is likely to occur if the risk algorithm is put into practice with unlabeled data. Four issues are addressed in the tables below. Projected costs for probation/parole supervision and demographic fairness are examined in Tables~\ref{tab:AllDist} through \ref{tab:NonwhiteDist}. Empirical coverage and its fairness are examined in Tables~\ref{tab:AllCover}  to \ref{tab:NonwhiteCover}. 

\subsection{Prediction Set Relative Frequency}

Table~\ref{tab:AllDist} shows the proportion of times in the forecast data each kind of prediction set occurs, conditional on the specified coverage probability $0.75$.\footnote
{
Because the confusion table indicated that the random forest classifier was often indecisive, some conformal prediction sets having more than one outcome class were anticipated. Consequently, greater precision was traded for less coverage. A coverage probability of $0.90$ was entertained, but not used.
}
Prediction sets with ``high'' alone are relatively rare: about 0.12 for all offenders. This addresses any concerns about supervisory overload for high-risk offenders. Apparently, there is no problem. The agency has the capacity to properly supervise about 15\% of offenders at intake who are classified as high risk.

\begin{table}[!htp]
\centering
\begin{tabular}{|c|c|c|c|c|c|c|}
\hline
H & M & L & H,M & H,L & M,L & Can't Tell \\
\hline
0.12 & 0.24 & 0.42 & 0.04 & 0.03 & 0.13 & 0.002 \\
\hline
\end{tabular}
\caption{All Offenders: Distribution of Mondrian Conformal Prediction Sets for the Forecasting Data: H = High, M = Moderate, L = Low, (N = 4700)}
\label{tab:AllDist}
\end{table}

\begin{table}[!htp]
\centering
\begin{tabular}{|c|c|c|c|c|c|c|}
\hline
H & M & L & H,M & H,L & M,L & Can't Tell \\
\hline
0.08 & 0.27 & 0.42 & 0.04 & 0.02 & 0.17 & 0.001 \\
\hline
\end{tabular}
\caption{White Offenders: Distribution of Mondrian Conformal Prediction Sets for the Forecasting Data: H = High, M = Moderate, L = Low, (N = 1471)}
\label{tab:WhiteDist}
\end{table}

\begin{table}[!htp]
\centering
\begin{tabular}{|c|c|c|c|c|c|c|}
\hline
H & M & L & H,M & H,L & M,L & Can't Tell \\
\hline
0.14 & 0.22 & 0.43 & 0.045 & 0.03 & 0.11 & 0.003 \\
\hline
\end{tabular}
\caption{Nonwhite Offenders: Distribution of Mondrian Conformal Prediction Sets for the Forecasting Data: H = High, M = Moderate, L = Low, (N = 3229)}
\label{tab:NonwhiteDist}
\end{table}

From the Tables~\ref{tab:WhiteDist} and \ref{tab:NonwhiteDist}, the proportion of forecasted high-risk offenders is 0.08 for Whites and 0.14 for Nonwhites, a difference of 0.06 ``favoring'' White offenders. Some might see this disparity as a fairness issue, but the re-arrest base rate for serious crimes is higher for Nonwhite offenders. One cannot expect a risk algorithm to correct for events over which it has no direct control \citep{Kleinberg2017}. Moreover, whether this disparity is large enough to matter needs to be addressed in the context of accuracy, which is considered shortly. 

The proportion of forecasted moderate-risk offenders overall is 0.24. The proportions are 0.27 for Whites and 0.22 for Nonwhites, a difference of 0.05. Some might see this disparity as ``favoring'' Nonwhites. But here too, fairness evaluations can be challenging. Arithmetic differences between White and Nonwhite offenders are almost the same for \{High\} and for \{Moderate\}, but the ratio is visibly larger for \{High\} because the proportion of \{High\} for both Whites and Nonwhites is so much smaller. This intimates another complication in the fairness controversies: Racial differences or racial ratios? 

By far the most common prediction set with a single outcome class is \{Low\}. Recall that \{Low\} is a forecast of no arrest and leads to almost no oversight by probation officers. For all offenders, and separately for White and Nonwhite offenders, that proportion is about .42. 

For all of the tables, the prediction sets with two elements are relatively rare. Two have useful information. \{Moderate, Low\} is the most common and conveys that the high-risk outcome class is unlikely. \{High, Moderate\} conveys that a re-arrest is likely but whether it is for a serious crime or not is undetermined.
 
Whether the \{High, Low\} prediction set is perplexing depends on whether the three outcome classes are treated as ordinal. They are conceptually ordinal by the yardstick of public safety, but they were defined as categorical in the analysis because the offenders forecasted in different outcome classes were not assumed to rank differently by some common crime etiology. Arguably, an offender who commits a homicide might be fundamentally different from an offender who commits a burglary, not worse on some shared scale of criminal inclinations. The same reasoning can be applied to offenders not re-arrested, compared to offenders who are.\footnote
{
These kinds of comparisons are complicated because every offender on probation or parole was arrested and sentenced, even those forecasted not to be re-arrested. This is a very selective collection of offenders much like the offenders on probation in this jurisdiction in previous years but not like the general adult population in the neighborhoods where they live.
}

No offenders in the holdout sample have all three outcome classes in their prediction set. This encouraging result is at least partly a product of the specified overall 0.75 coverage probability. Somewhat more uncertainty was traded for somewhat more precision.

The ``Can't Tell'' class represents forecasting attempts that fail; there is no forecast at all. There are at least two possible causes, but these are not the result of errors in the conformal procedures. Sometimes the classifier passes along three risk scores that are extremely similar. And because the three Mondrian distributions are employed independently, one can on occasion get results that imply a sum greater than $1.0$. In any case, the ``Can't Tell'' prediction set is extremely rare. By this yardstick, the risk assessment algorithm is performing quite well.

In summary, some prediction sets are far more common than others. For the prediction sets to be useful for decisions about probationers, prediction sets containing a single outcome class should dominate. In Table~\ref{tab:AllDist}, singleton prediction sets account for about three quarters of all realized prediction sets. Whether that is good enough is a matter for stakeholders to decide. There is some evidence of demographic unfairness, but the racial differences are modest with White offenders sometimes ``favored'' and Nonwhite offenders sometimes ``favored.'' This is also consistent with racial differences in base rates over which departments of probation or parole have no direct control. The racial proportions for \{Low\} are virtually identical. This is important because probation ``lite'' is a preferred probation sentence.

\subsection{Set Accuracy as Empirical Coverage}

An evaluation of forecasting correctness requires distinguishing between precision, coverage, and accuracy. Precision is determined by the number of outcome classes in a prediction set. Sets with fewer outcome classes are more precise. Coverage is the probability that when a forecast is made, it is correct. This tradeoff between precision and coverage must be considered before the data analysis begins. Coverage is a theoretical construct that is specified before the data analysis begins. That is when the tradeoff between precision and coverage must be considered. Coverage is the target for empirical accuracy.\footnote
{
Formally, the marginal coverage of a prediction set $\widehat{C}_n$ at level $1-\alpha$ is
\[
    \mathbb{P}\big( Y_{n+1} \in \widehat{C}_n(X_{n+1}) \big) \geq 1 - \alpha,
\]
where $\alpha$ is the miscoverage rate, and the probability is taken over the joint distribution of the calibration data and the test point $(X_{n+1}, Y_{n+1})$. Specified coverage is \emph{at least} $1 - \alpha$. For this analysis, the miscoverage rate was specified as $0.25$ so that the specified coverage is $1 - 0.25 = 0.75.$
\label{fn:coverage}
}
Empirical accuracy, also called empirical coverage, is determined in a holdout sample as the proportion of correct forecasts made by a prediction set. Ideally, theoretical coverage and empirical coverage should be about the same. 

Tables~\ref{tab:AllCover} to \ref{tab:NonwhiteCover} show the results. The empirical coverage is at or above the theoretical coverage of 0.75, indicating the conformal prediction approach taken likely is working as intended. But there are substantial differences in accuracy depending on the content of a prediction set. In general, prediction sets containing a single outcome class will be more precise but less accurate than prediction sets containing more than a single outcome class. For example, \{Moderate\} would likely be more precise and less accurate than \{High, Moderate\}. This is in the nature of the conformal forecasting mathematics and not special to any given data set or classifier. 

For stakeholders especially concerned about public safety, the high-risk outcome class is often the top priority. But forecasting accuracy for high-risk re-arrests is disappointing. Once again, the classifier's difficulties distinguishing high-risk offenders from moderate-risk offenders cascade through the analysis.

For Tables~\ref{tab:AllCover} to \ref{tab:NonwhiteCover}, forecasts of re-arrest for high-risk offenses have an accuracy of about .45; a high-risk forecast will be correct a little less than half the time. This is disappointing, but it is a meaningful improvement over the prediction success reported by the confusion table. The Mondrian conformal forecasts for high-risk offenders are correct about 45 times in 100 offenders, while the confusion table prediction success rate for high-risk offenders is about 37 times out of 100 offenders; the former is about 22\% larger. Whether this is large enough to matter is for stakeholders to decide, but there are good reasons why one should expect greater accuracy for Mondrian conformal prediction sets. 

The discrepancy between prediction accuracy for the high-risk class reported in the confusion table and the empirical forecasting accuracy of singleton high-risk conformal prediction sets is attributable to a fundamental difference in the decision rule underlying each quantity, not to any inconsistency in the conformal procedure itself. The reason can be found in the competing decision rules.

The confusion table is constructed from the classifier's \emph{argmax} rule: a case is labeled ``Predicted High'' whenever the estimated probability of the high-risk class exceeds those of the low-risk and moderate-risk classes, regardless of the absolute magnitude of that probability or its difference compared to the probabilities of alternative classes. Because the random forest was trained on data in which the high-risk class was oversampled to equalize classification error rates, its probability estimates for the high-risk class are systematically inflated. This inflation is often sufficient to let predictions of high risk narrowly win the three-way comparison even when the underlying evidence for high risk is weak. This is precisely what produces the elevated false-discovery rate (0.63) observed for the Predicted High column.

The construction of a Mondrian conformal singleton set imposes a distinctly different, and considerably more stringent, requirement. Rather than a single relative comparison between three classes, a singleton {High} prediction set requires that three separately calibrated conditions hold simultaneously. The conformity score for {High} must clear its own threshold (calibrated against the empirical score distribution of true high-risk cases), \emph{and} the conformity scores for both {Low} and {Moderate} must fail to clear their respective thresholds. In many cases, High wins the \emph{argmax} comparison—it is the classifier's top choice—yet the exclusion conditions for Low and Moderate are not met: their conformity scores fall just short of being ruled out, so they remain in the prediction set alongside High. Such cases are consequently assigned to a non-singleton prediction set rather than to {High} alone. The higher empirical accuracy observed for singleton conformal predictions, therefore, does not indicate that the conformal calibration step corrects the classifier's probability estimates in any absolute sense. Rather, it capitalizes on the empirical fact that a \emph{singleton} is substantially rarer and more selective than an \emph{argmax} victory.

The forecasting improvement with Mondrian conformal prediction sets is, therefore, best understood as a precision--selectivity tradeoff inherent to set-valued prediction, not as evidence that the coverage target of 0.75 functions as a stricter classification threshold. The major consequence for practice is that \emph{the Mondrian approach yields unambiguous outcome class forecasts that are more reliable.}

\begin{table}[!htp]
\centering
\begin{tabular}{|c|c|c|c|c|c|c|}
\hline
H & M & L & H,M & H,L & M,L & Can't Tell \\
\hline
0.45 & 0.66 & 0.88 & 0.68 & 0.86 & 0.97 & 0.00 \\
\hline
\end{tabular}
\caption{All Offenders: Coverage for Conformal Prediction Sets for the Forecasting Data: H = High, M = Moderate, L = Low, Overall Coverage = 0.77 (N = 4700)}
\label{tab:AllCover}
\end{table}

\begin{table}[!htp]
\centering
\begin{tabular}{|c|c|c|c|c|c|c|}
\hline
H & M & L & H,M & H,L & M,L & Can't Tell \\
\hline
0.47 & 0.70 & 0.89 & 0.67 & 0.91 & 0.98 & 0.001 \\
\hline
\end{tabular}
\caption{White Offenders: Coverage for Conformal Prediction Sets for the Forecasting Data: H = High, M = Moderate, L = Low, Overall Coverage = 0.81 (N = 1471)}
\label{tab:WhiteCover}
\end{table}

\begin{table}[!htp]
\centering
\begin{tabular}{|c|c|c|c|c|c|c|}
\hline
H & M & L & H,M & H,L & M,L & Can't Tell \\
\hline
0.44 & 0.64 & 0.89 & 0.69 & 0.85 & 0.96 & 0.00 \\
\hline
\end{tabular}
\caption{Nonwhite Offenders: Coverage for Conformal Prediction Sets for the Forecasting Data: H = High, M = Moderate, L = Low, Overall Coverage = 0.75 (N = 3229)}
\label{tab:NonwhiteCover}
\end{table}

 All of the prediction sets that are not \{High\} have empirical coverage near the theoretical coverage of 0.75. Prediction sets that contain \{Low\} have substantially better empirical coverage. This is not an error in conformal methods, but fully consistent with the formal definition of coverage in footnote~\ref{fn:coverage} in which the probability that the true future outcome class falls in a prediction set can be equal to \emph{or greater than} the specified coverage. Looking back at the random forest confusion table, each of \{High\}, \{Moderate\}, or \{Low\} is at least a bit more accurate than the corresponding predicted classes. Note that when \{Low\} is the conformal forecast, it is correct almost 90\% of the time in all three tables. The confusion table does nearly as well. Random forest seems quite able to find offenders unlikely to re-offend.

In short, forecasts from Mondrian conformal prediction sets can have better empirical coverage and, therefore, more reliability than predictions extracted from a confusion table. Perhaps more important, Mondrian conformal prediction sets provide valid measures of uncertainty in finite samples (i.e., not asymptotically) conditional on the coverage probability specified, and highlight the tradeoffs between coverage and precision. These benefits accrue with no assumptions about the distributional form of the forecasts or of the joint probability distribution responsible for the data. Exchangeability is enough.

Finally, racial unfairness appears not to be an issue for forecasting accuracy. White offenders and Nonwhite offenders have very similar empirical coverage estimates across all of the prediction sets. Moreover, even though there is a larger number of high-risk forecasts for Nonwhites than Whites, there is essentially no difference in forecasting accuracy; \emph{the larger number of high-risk forecasts is not explained by false discoveries}. In short, there is no evidence of forecasting accuracy unfairness for Nonwhite offenders compared to White offenders.

\section{Some Policy Implications}\label{sec7}

There are recent documents that try to provide practical guidelines for the use of AI in criminal justice \citep{CouncilOnCriminalJustice2026}. These can be quite useful as a list of important challenges to address, but they can be misread as prescriptions for particular settings that typically vary in resources, decisions to be affected, AI tools to be used, legal constraints, political landscape and technical expertise. In contrast, the recommendations in this section should generalize quite well, largely because they center on technical matters well addressed in the statistical and computer science literature.  

Conformal prediction sets, not the output of the random forest classifier, should be used to make real decisions one-by-one about real offenders. Relying on the classifier output alone risks introducing the equivalent of coin flips into the forecasting process. This applies to any of the popular classifiers including logistic regression, gradient boosting, random forests, and neural networks (deep or shallow). 

Improvements in forecast reliability can vary in size depending on how well the classifier performs. When a classifier makes clear distinctions between outcome classes, reliability is already high. Then, the main benefit of Mondrian conformal prediction sets is valid forecasting uncertainty with no need for additional assumptions or asymptotics. There is nothing comparable from classifiers or confusion tables by themselves. Even when there are very modest improvements in reliability, the manner in which Mondrian conformal prediction sets represent uncertainty is well worth the extra steps.

Another benefit is better transparency. It seems self-evident that offenders have a right to know how their risk class was determined. Importance plots and partial dependence plots can be very instructive. Valid uncertainty assessment should be available as well. What was the content of the offender's prediction set, what was the coverage probability, and what was done if the prediction set contained more than one outcome class? Such information should also be of interest to an offender's probation officer and, when aggregated over offenders, to agency administrators.

A variety of decision aids can be provided when the prediction set forecasts more than a single outcome class. For example, one can as a policy matter choose the class representing the greatest threat to public safety. Hence if the prediction set is \{Moderate, Low\}, choose Moderate. If there are concerns that the risk assessment tool is ``overpredicting'' the High and Moderate outcome classes, choose Low, which is the outcome class representing the lesser threat to public safety. 

When permitted by local administrative regulations and criminal law statutes, it can be useful to have procedures in place that correct apparent forecasting errors. For example, there can be ways to ``step down'' or ``step up'' an assigned class if an offender's behavior while under supervision merits a change. Alternatively, there might be routine re-evaluations of all offenders every 3 months after which changes in supervision can be made. None of these interventions are necessarily informed by updated risk assessments, although it would be possible with the right data to develop rigorous forecasts for offenders ``in progress.'' 

\section{Conclusions}\label{sec8}

Mondrian conformal prediction sets can improve forecast reliability and provide valid inference about forecasting uncertainty in finite samples when conformity scores are exchangeable. The data need not be realized in an iid manner and no functional forms are assumed for the distribution of the forecasts. There is no apparent downside compared to confusion tables from classifiers. At the very least, when a classifier is indecisive for a given case, a response of ``don't know'' is more honest than the equivalent of a surreptitious coin flip.

\section*{Appendix on Mondrian Conformal Prediction Sets}

In this paper, the split sample approach informs the construction of three random, disjoint subsets of the data: a training subset, a calibration subset, and a forecast subset. The training data are used with the random forest classifier to produce fitted class probabilities. The calibration data are used to operationalize the calculations required for the Mondrian conformal prediction sets. The forecast data are used to evaluate how well those prediction sets empirically perform. Once the risk assessment procedure is put into practice, the forecast data are replaced by unlabeled cases that need their labels forecasted.

The random forest algorithm, or any classifier, is used as one ordinarily would with a categorical response variable. This is business as usual for a wide variety of applications. The remaining steps are standard for conformal prediction sets with a few differences because of the Mondrian methods used.

The calibration data is able to show how well the classifier performs because the true outcome class is known for every case. The predict function from the random forest fit to the training data is applied to the predictors in the calibration data. The output is an $N_c$ by $K$ array of predicted class probabilities that for this analysis are risk scores. These risk scores are distinct from the conformity scores but are close precursors.

Suppose for a single case the calibration risk scores are 0.10 for high risk, 0.20 for moderate risk, and 0.70 for low risk. If the actual outcome class is high risk, the conformity score is 0.10; the case conforms poorly to the truth. If instead the actual outcome class were moderate risk, the conformity score would be 0.20; the case conforms poorly to the moderate risk class. If instead the actual outcome class were low risk, the case would conform much better, with a conformity score of 0.70.

These three suppositions illustrate a general mapping. Given a case's risk scores, any hypothesized outcome class has a corresponding conformity score. For a calibration case, only one of these risk scores is realized because the true outcome class is known; the conformity score is for the actual class. That single value, for every case in the calibration data, is what determines the class-specific thresholds described below. The same mapping becomes useful again at the forecasting stage, when the true class is unknown. In that setting, each of the three hypothesized classes yields its own conformity score, and all three are checked against the calibration thresholds to build the prediction set.

For the Mondrian approach used here, the conformity scores are organized into three disjoint \emph{subsets} according to each case's \emph{actual} outcome class. There are $N_{c,h}$ conformity scores for calibration cases whose true outcome is high risk. There are $N_{c,m}$ conformity scores for calibration cases whose true outcome is moderate risk. There are $N_{c,l}$ conformity scores, one for each calibration case whose true outcome is low risk.

These three counts need not be equal because the calibration data typically do not divide evenly across outcome classes. For each outcome class, the corresponding subset of conformity scores gives a distribution of conformity scores conditional on the calibration data and the trained classifier, both of which are treated as fixed. Ideally, the scores within each subset span a wide range, from near 1.0 (cases the classifier fits well) down toward 0.0 (cases the classifier fits poorly), reflecting genuine variation in how well the classifier performs within that class. The three distributions show how well the classifier performs on a holdout sample \emph{when the true outcome class is known}. As such, they provide the basis for the thresholds used below, when forecasts are needed for unlabeled cases whose true outcome is \emph{not known}. What one learns with the true outcome class is known is carried over when the true outcome class is not known.

But how good is a good conformity score? Suppose, before the data analysis begins, a decision is made that there is good enough conformity for the largest 90\% of the conformity scores within each of the three conformity score subsets.\footnote
{
This is a decision that should be made by the ultimate users of the risk instrument. It is a policy decision, although it has statistical consequences. 
}
The 90\% serves as a ``coverage probability'' of 0.90. An important implication is that the lower bound of the top 90\% is, approximately, the score at the 0.10 quantile of the low-risk subset's distribution. That conformity score value might be 0.55; call it $\hat q_l$. In practice the quantile used is not exactly the 0.10 quantile, but a very slightly higher one, calculated as roughly $\lceil (N_l+1) \times 0.10 \rceil / N_l$ rather than $N_l \times 0.10$. This finite-sample correction is what allows split conformal to guarantee coverage of at least 90\%, rather than only approximately 90\%, even when a fixed overall $N$ is not large.\footnote
{
For the exact formula and its derivation, see \citep{Angelopoulos2023}. Coverage is analogous to the confidence level in conventional statistics.
}

The same reasoning and coverage probability are used with the other two outcome class subsets, yielding thresholds $\hat q_h$ and $\hat q_m$ for the high and moderate outcome classes, computed the same way but using $N_h$ and $N_m$ in place of $N_l$; the resulting quantile values for the lower bound will likely differ across the three classes, both because the three score distributions differ and because $N_h$, $N_m$, and $N_l$ are generally unequal.\footnote
{
Other kinds of split sample approaches do not subset the calibration data by outcome class \citep{Angelopoulos2025}. The Mondrian method trades smaller subset sizes for the ability to calibrate each outcome class separately. This can be useful when some outcome classes are more important than others. The subset size does not enter the quantile calculation as an explicit parameter, but a smaller subset still performs as it should but in a more coarse-grained manner.
}

Consider an $N+1$ unlabeled case needing a forecast among the same three outcome classes. Because the truth is not known, a thought experiment follows. The case's predictor values are used as input to the trained random forest's predict function, which outputs a risk score for each outcome class. For low risk, the risk score might be 0.70.

Now, hypothetically, suppose that low risk is the true, future outcome class for this case. Under that supposition, the 0.70 becomes a conformity score for the $N+1$ case with respect to low risk, and it can be compared to $\hat q_l$, the threshold built from the low-risk subset of the calibration data. If 0.70 exceeds $\hat q_l$, the low-risk class is included in the case's prediction set.

The same procedure is repeated for the other two outcome classes, each compared against its own threshold. The risk score the classifier assigns to high risk is treated as a hypothetical conformity score compared to $\hat q_h$. The risk score the classifier assigns to moderate risk is treated as a hypothetical conformity score and compared to $\hat q_m$. Any class whose hypothetical conformity score exceeds its own threshold is included in the prediction set. Classes that do not clear their threshold are excluded. The prediction set for the $N+1$ case can therefore end up containing one class, more than one class, all three classes, or because nothing guarantees at least one class clears its threshold, none at all.

The coverage guarantee applies to the procedure, not to any single case's forecast. It says that if this rule were applied to many cases whose true outcome really is low risk, for instance, about 90\% of those cases would have low risk included in their prediction set. The same reasoning holds separately for high-risk cases with respect to $\hat q_h$, and for moderate-risk cases with respect to $\hat q_m$. It is not a claim that any individual forecast is correct with probability 0.90 \citep{Barber2021}.

But for that claim to be credible, a strong assumption is required: the conformity scores must be exchangeable. It helps to be precise about what object this assumption addresses. Unlike the risk scores discussed earlier, which here form an $N \times 3$ array (one score per case, per outcome class), the low-risk subset's conformity scores form a single column: one conformity score per case, for the $N_l$ cases whose true outcome is low risk. Exchangeability is a property of this single sequence of scores: it requires that the joint distribution of the sequence is unchanged by any reordering of the cases that produced it. For example, suppose the rows of the data frame holding these $N_l$ conformity scores are reordered, cases and their scores moving together. Exchangeability says the distribution of that sequence of scores is exactly the same regardless of the order in which the rows happen to appear. 

An important feature of exchangeability is that conformal inference relies on the \emph{rank} of the future score within the calibration conformity score distribution. For example, in a subset of 100 calibration scores, a quantile value of 0.44 means that 44\% of those scores fall below it --- roughly the 44th score when the scores are arranged low to high. Exchangeability guarantees this rank is uniformly distributed, which is what delivers the provable finite-sample coverage guarantee.\footnote
{
Exchangeability means the joint distribution of $S_1, \dots, S_{N_l+1}$ is unchanged under any permutation, so no score is statistically distinguishable from any other. By this symmetry, $S_{N_l+1}$ is equally likely to occupy any of the $N_l+1$ rank positions when the scores are sorted low to high; its rank is uniformly distributed on $\{1, \dots, N_l+1\}$. This is what guarantees that the $\lceil(1-\alpha)(N_l+1)\rceil$-th smallest calibration score is a valid $(1-\alpha)$ threshold for the future score. The same argument applies separately within the high-risk and moderate-risk subsets, using $N_h$ and $N_m$ in place of $N_l$.
}
If the scores were not exchangeable, the future score's rank need not be uniform. Its distribution could be shifted or skewed relative to the calibration scores, and the calibration quantile would no longer be a valid threshold.

Another useful feature of exchangeability is that it does not require the data to be independent and identically distributed, only that their joint distribution be invariant to permutation. This matters because the $N+1$ case materializes only after the training and calibration data exist; under exchangeability, that ordering in time does not affect the validity of the guarantee.\footnote
{
The Mondrian formulation (named after the French artist) is easily misunderstood. One conditions on features of the data. \citet{Barber2021} prove that conditioning on subsets of cases based on anything \emph{computed from the training data} (e.g., estimated outcome classes) will violate finite-sample claims of valid coverage. Note that the actual outcome class for these data came with the calibration data; it was not estimated from it.
}

For the analyses reported above, the case for exchangeable conformity scores is strong \citep{Lei2018}. In split conformal inference, both the calibration sample and the trained classifier are treated as fixed: the classifier is estimated once and not re-fit, and the calibration data are a fixed, already-observed batch. Fixing the classifier does one specific piece of work: because a fixed function that is applied identically and independently to each case preserves exchangeability, any conformity scores for which the trained random forest is responsible are exchangeable. Fixing the classifier therefore transmits exchangeability. It does not create it.

\backmatter

\section*{Declarations}
\begin{itemize}
\item Funding: Not applicable
\item Conflict of interest: Not applicable
\item Data availability: This is a matter for the agency providing the data to decide.
\item Code availability: Yes in principle. The several thousand lines of code is available in R. But the code tackles several very different tasks some of which may be of little interest. Comments in the code itself is the only code documentation available, and it is very specific to the data used in this paper.   
\end{itemize}

\bibliography{ADPP}
\end{document}